\documentclass[showpacs,preprintnumbers,twocolumn,superscriptaddress,floatfix]{revtex4}
\usepackage{amsmath}
\usepackage{amsfonts}
\usepackage{amssymb}
\usepackage{graphicx}
\usepackage{subfigure}

\begin{document}

\title{$\langle p_\text{t} \rangle$ Systematics and $m_\text{t}$--Scaling}
\author{Richard Witt}
\email[]{witt@bnl.gov}
\affiliation{LHEP, University of Bern, Switzerland}
\collaboration{for the STAR Collaboration}

\date{\today}

\begin{abstract}
An enhancement in the number of strange particles produced in relativistic heavy ion collisions is
expected to coincide with the formation of a deconfined state  of partonic matter\cite{Rafelski82}.
Measurements of transverse momentum spectra for strange  particles emerging from $p+p$ collisions are
used as a baseline to which similar measurements from heavy ion collisions are compared.  In addition,
several observations from $p+p$ collisions, such as the variation of $\langle p_\text{t} \rangle$ with
particle mass and with event multiplicity, are interesting in their own right.  We present
measurements of the transverse momentum spectra and $\langle p_\text{t} \rangle$ systematics for
strange and non-strange particles from $p+p$ collisions at $\sqrt{s}$=200 GeV.  We show the
dependence of the $\langle p_\text{t} \rangle$ on measured charged multiplicity and on particle mass.
We will also demonstrate the ability to scale the transverse mass spectra of various species onto a
single universal curve for our $p+p$ data (an effect known as $m_\text{t}$--scaling) and the failure of
this scaling when applied to our Au+Au data.  The work presented here was presented as a poster at Quark Matter 2004.
\end{abstract}

\maketitle

\section{Introduction \label{intro}}
Measurements on $p+p$ colliding systems at RHIC provide a crucial baseline measurement with which to
compare measurements from Au+Au collisions.  These comparisons offer a means to distinguish nuclear
medium effects from what might be expected in a superposition of $A$ nucleon+nucleon collisions.

However, $p+p$ systems are interesting in their own right.  Previous measurements have shown a
systematic dependence of the $\langle p_\text{t} \rangle$ on particle mass as well as event
multiplicity.  Measurements from UA1\cite{Bocquet96} of the $\langle p_\text{t} \rangle$ of $\Lambda$, $\text{K}^{0}_{s}$, and charged hadrons as a function of the $N_\text{ch}/\eta$ from $p+\bar{p}$ collisions at $\sqrt{s}$=630 GeV show a clear increasing trend with $N_\text{ch}/\eta$ and this trend is stronger for the $\Lambda$ than for the $\text{K}^{0}_{s}$ or the pion-dominated charged inclusives.

Lower energy ISR data ($\sqrt{s}\sim$23 GeV -- 63 GeV)\cite{BSCref1} seem to also show a mass dependence for $\langle p_\text{t} \rangle$ which has been parameterized\cite{ISRparam} with the form $0.7*m^{0.4}$.  An increase in the $\langle p_\text{t} \rangle$ with particle mass should also be reflected in the inverse-slope parameter.  Model calculations that include the  effects of ``mini-jets" (multi-parton interactions) show such a dependence of the inverse-slope parameter on particle mass\cite{Dumitru99} while hydro calculations show only a weak dependence on mass at best.

Identified particle spectra ($\pi^{\pm}$, K$^{\pm}$, p \& $\bar{\text{p}}$) at ISR energies have been  successfully described\cite{Gatoff92} by a ``universal" parameterization of the form shown in equation \ref{equ:univ}

\begin{equation}
E\dfrac{d\sigma}{d^{3}p} = A\dfrac{e^{-\text{m}_\text{t}/T}}{\text{m}^{\lambda}_\text{t}}
\label{equ:univ}
\end{equation}

where $m_\text{t}$ is the transverse mass ($\sqrt{p_\text{t}^{2}+m^2}$).  This phenomenon is known as
$m_\text{t}$--scaling and would seem to suggest that over a given range of $m_\text{t}$, spectra of particles with different masses would have similar slopes.  Also, even if the spectra have no relative scaling factors applied, the yields will be similar at a given $m_\text{t}$.  It has been suggested that in Au+Au collisions at sufficiently high energy the parton density increases to the point of saturation, forming a state referred to as the ``Color Glass Condensate" (CGC)\cite{McLerran94,McLerran01}.  The intrinsic momentum scale of the CGC allows a single function to describe produced particle spectra over a range of energy, centrality, and system size.  This idea has been tested on data from $p+p$ at the CERN SPS and, more recently, on minimum bias $\sqrt\text{s} = 130$ GeV Au+Au data from the PHENIX collaboration at RHIC\cite{Bielich02,BielichXXX}.

\section{Experiment \label{exp}}
STAR is a multi-system experiment the heart of which is the large volume Time Projection Chamber (TPC).  The TPC provides excellent tracking of charged particles with good momentum resolution (~2-7\% for $\pi$'s out to a $p_\text{t}$ of ~4 GeV/$c$) over a broad rapidity window (~1.8 units in $\eta$).  Particle identification is done via $dE/dx$ and, more recently, with a Time of Flight (TOF) system\cite{TOF}.  The full details of the STAR experiment subsystems are available elsewhere\cite{NIM}.

Identification of strange particle decays in STAR is done via a topological algorithm in which several geometrical cuts are applied to the charged particle tracks.  Once identification is done and the parent momenta determined, the resulting spectra must be corrected for finite detector acceptance and efficiency.  For this we use a technique known as ``embedding".

In embedding, Monte-Carlo generated tracks are propagated through a full GEANT\cite{GEANT} simulation of the STAR detector response and then added at the pixel level to signals from real data events.  The embedded event is then processed with our standard event reconstruction software and a search is made to see if the embedded particle's track is reconstructed.  In this way we obtain correction factors which account for both efficiency and acceptance losses, as a function of $p_\text{t}$, in a single procedure.


\section{Results \label{results}}
\subsection{$p_\text{t}$ Spectra \label{spectra}}
STAR has measured the transverse momentum spectra for particles with masses up to and including the $\Lambda(1520)$ baryon in $p+p$ systems with $\sqrt{s}=200$ GeV.  Present statistics are insufficient for an $\Omega^{-}$ spectrum in $p+p$.  The $p_\text{t}$ ranges of the measured spectra depend on the particle species due to the differing means of particle identification.  Pion, kaon, and proton spectra from the TPC have the greatest precision at low $p_\text{t}$ but are the most limited in extent (out to $\sim$1 GeV/$c$ in total momentum) due to the overlapping of the Bethe-Bloch bands.  The TOF increases the upper limit to about 3 GeV/$c$.  Strange particles' decays can be reconstructed over a wide range of transverse momentum $0.5~\text{GeV}/c \lesssim p_\text{t} \lesssim 30 ~\text{GeV}/c$ but are limited by the available statistics.

\begin{figure}[h]
\includegraphics[width=6.5cm]{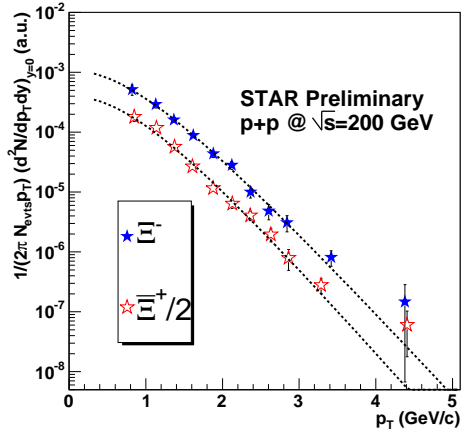}
\caption{Mid-rapidity transverse momentum spectra for $\Xi^{-}$ and $\overline{\Xi}^{+}$ from $p+p$ at $\sqrt{s}=200$ GeV, $\left|y\right|<0.75$.  Error bars are statistical only}
\label{fig:ppXi}
\end{figure}

Figure \ref{fig:ppXi} shows the mid-rapidity $p_\text{t}$ spectra for $\Xi$ and Anti-$\Xi$ baryons.
The fits are exponentials in $m_\text{t}$ of the form:

\begin{equation}
\dfrac{1}{2\pi N_\text{evts} p_\text{t}}\dfrac{d^{2}N}{dp_\text{t}dy} = Ae^{-m_\text{t}/T}
\label{equ:mtfit}
\end{equation}

which describes the data reasonably well at low $p_\text{t}$.  But we already begin to see the deviation of the power-law-like tail above $\sim$4 GeV/$c$.  While the statistics do not allow binning in event multiplicity for the $\Xi$'s, the situation is quite different for the $\text{K}^{0}_{s}$ and $\Lambda$\cite{JohnAndMark}.


\subsection{$\langle p_\text{t} \rangle$ vs. Multiplicity and Mass \label{meanpt}}
From the fits to the $p_\text{t}$ spectra we may determine the $\langle p_\text{t} \rangle$ in each multiplicity bin.  The results of this determination can be seen in the figure \ref{fig:lamk0meanpt} below.  Note the suppressed zero on the vertical scale.
\begin{figure}[h]
\includegraphics[width=6.5cm]{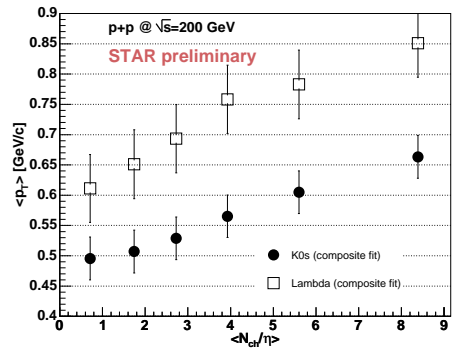}
\caption{Dependence of $\Lambda$ and $\text{K}^{0}_{s}~\langle p_\text{t} \rangle$ on uncorrected $\langle N_\text{ch} \rangle/\eta$, $\left|y\right|<0.5$.  Error bars are statistical only.}
\label{fig:lamk0meanpt}
\end{figure}
Both species show a clear rising trend with uncorrected $\langle N_\text{ch} \rangle/\eta$, though the trend seems to be slightly stronger for the $\Lambda$'s.  Both $\text{K}^{0}_{s}$ and $\Lambda$ production seems to be biased more toward the harder collisions that are indicated by the higher uncorrected $\langle N_\text{ch} \rangle/\eta$.

\begin{figure}[h]
\includegraphics[width=6.5cm]{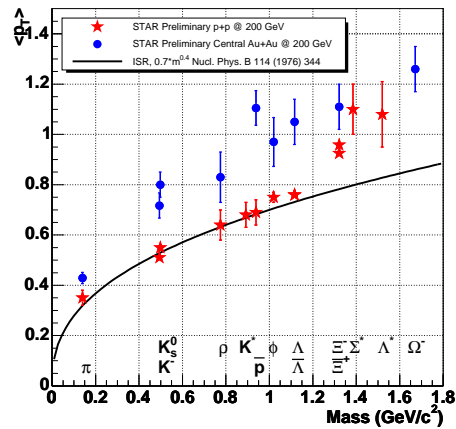}
\caption{$\langle p_\text{t} \rangle$ vs. particle mass for $p+p$ at $\sqrt{s}=200$ GeV and Au+Au at $\sqrt{s_\text{NN}}=200$ GeV.}
\label{fig:meanptvsmass}
\end{figure}

We present in the figure \ref{fig:meanptvsmass} the trend of the $\langle p_\text{t} \rangle$ with particle mass for the 200 GeV $p+p$ and central Au+Au systems.
The black line is from the $0.7*m^{0.4}$ parameterization of the ISR data which included masses only up to the $\Lambda$.  The difference between the Au+Au and $p+p$ $\langle p_\text{t} \rangle$ for each particle is believed to be due mostly to the presence of radial flow\cite{STAR03} in the Au+Au system.  This effect is weaker for the more massive particles.  Although the ISR parameterization works well for the lower mass particles despite the large difference in beam energy, the deviations become significant for the higher masses where the increase in $\langle p_\text{t} \rangle$ is nearly linear with mass.

\subsection{$m_\text{t}$--Scaling in $p+p$ \label{ppmtscaling}}
\begin{figure}[h]
\includegraphics[width=6.5cm]{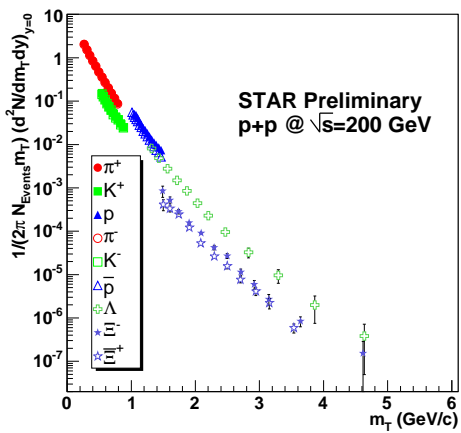}
\caption{Transverse mass spectra from $\sqrt{s}=200$ GeV $p+p$ collisions with no relative scaling factors applied.  Error bars are statistical only.}
\label{fig:ppnoscale}
\end{figure}

\begin{figure}[h]
\subfigure[1][Arbitrarily scaled spectra]
  {
  \includegraphics[width=6.5cm]{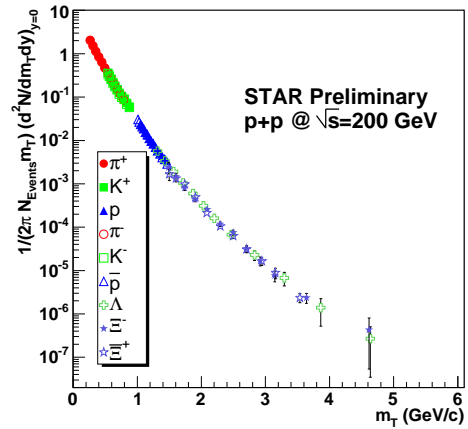}
  \label{fig:pplowptscaled}
  }
\subfigure[2][Ratio of data to overall fit]
  {
  \includegraphics[width=6.5cm]{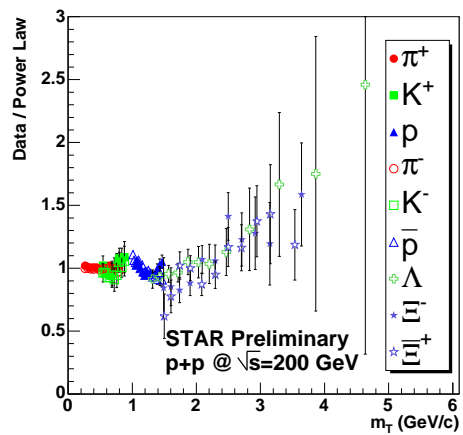}
  \label{fig:ppmtratio}
  }
\caption{Transverse mass spectra from $\sqrt{s}=200$ GeV $p+p$.  Error bars are statistical only.}
\end{figure}

Figure \ref{fig:ppnoscale} shows the spectra with no scaling factors applied.  The difference in yields at any given $m_\text{t}$ shows that $m_\text{t}$--scaling does not work in the absolute sense and can only be approximate at best for our $p+p$ data.  Taking the $m_\text{t}$ spectra for the various species from the $p+p$ and scaling them by arbitrary factors it seems likely that they would be describable by a single function.
The spectra in figure \ref{fig:pplowptscaled} have been scaled by arbitrary factors to achieve the best overlap and were fit with a power law of the form:
\begin{equation}
\dfrac{1}{2\pi N_\text{evt} m_\text{t}}\dfrac{d^{2}N}{dm_\text{t}dy} = \dfrac{A}{\left[1+\dfrac{m_\text{t}}{B}\right]^n}
\label{equ:pwrlaw}
\end{equation}
Figure \ref{fig:ppmtratio} is the point-by-point ratio of the data to the fit.  The fit is good to within 20\% out to almost 3 GeV/$c$.

\begin{figure}[h]
\includegraphics[width=6.5cm]{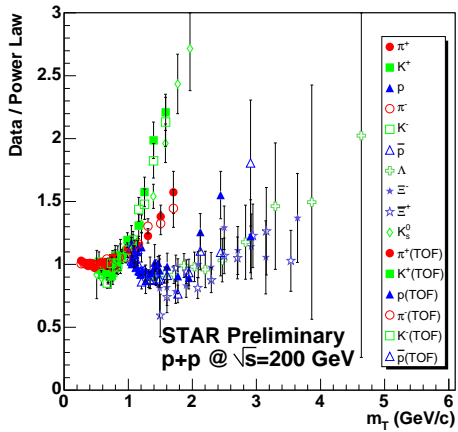}
\caption{Ratio of data to overall fit for scaled transverse mass spectra from $\sqrt{s}=200$ GeV $p+p$ collisions including Time-Of-Flight and $\text{K}^{0}_{s}$.  The TOF spectra provide a greater range in $p_\text{t}$ than those from $dE/dx$ alone.  Error bars are statistical only.}
\label{fig:ratiowithTOF}
\end{figure}

Adding in the data from the TOF as well as the $\text{K}^{0}_{s}$ data, but retaining the fit to the spectra from figure \ref{fig:pplowptscaled} we see in figure \ref{fig:ratiowithTOF} that the meson spectra trend away from the apparent curve established by the previous fit.  The trend becomes apparent due to the extended $p_\text{t}$ range availible in the TOF and $\text{K}^{0}_{s}$ spectra.
The onset of deviation from the original fit at $m_\text{t}\approx$1 GeV/$c$ is interesting in that the hard scatterings are beginning to become an important contribution to the spectra in this region.  The baryons seem to maintain agreement further out in $m_\text{t}$ and the delayed departure of the pions could be due to contributions feeding-down from higher mass decays.

\subsection{$m_\text{t}$--Scaling in Au+Au \label{AuAumtscaling}}
\begin{figure}[h]
\includegraphics[width=6.5cm]{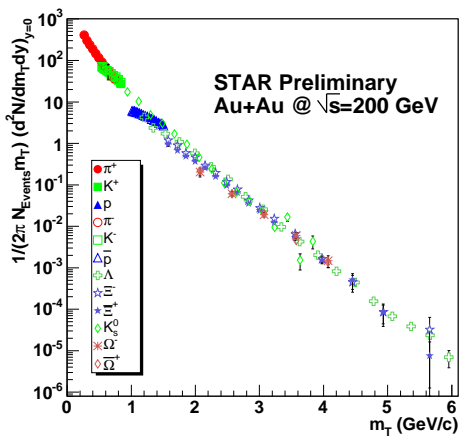}
\caption{Arbitrarily scaled transverse mass spectra from $\sqrt{s_\text{NN}}=200$ GeV Au+Au collisions.  Error bars are statistical only.}
\label{fig:scaledAuAu}
\end{figure}

When scaling factors are applied to the Au+Au $m_\text{t}$ spectra no absolute scaling is evident and only very little overlap can be achieved.  Figure \ref{fig:scaledAuAu} shows the scaled Au+Au spectra.
Deviations from the universal fit in $p+p$ and Au+Au are very different in nature.  In Au+Au the likely presence of radial flow changes the spectral shapes at low-$m_\text{t}$ eliminating the universal curve that was present in the $p+p$ spectra.  The Au+Au spectra are also different in shape, being flatter than those from p+p.

\section{Conclusions and Outlook \label{conclusions}}
We have presented preliminary results from analyses of the $p_\text{t}$ spectra of $\pi^{\pm}$,  $\text{K}^{\pm}$, p, $\bar{\text{p}}$, $\text{K}^{0}_{s}$, $\Lambda$, and $\Xi$ from $\sqrt{s}=200$ GeV $p+p$ collisions and $\pi^{\pm}$, $\text{K}^{\pm}$, p, $\bar{\text{p}}$, $\text{K}^{0}_{s}$, $\Lambda$, $\Xi$, and $\Omega$ from $\sqrt{s_\text{NN}}=200$ GeV Au+Au collisions measured by the STAR collaboration at RHIC.

The measured $\langle p_\text{t} \rangle$ for $\text{K}^{0}_{s}$ and $\Lambda$ show a clear dependence on the uncorrected average charged particle multiplicity per unit $\eta$ in the $p+p$ collisions.  The dependence is slightly stronger for the $\Lambda~\langle p_\text{t} \rangle$ than for the $\text{K}^{0}_{s}~\langle p_\text{t} \rangle$, while production of both species seems to be more probable in the harder, higher multiplicity collisions.  The mass dependence of the $\langle p_\text{t} \rangle$ agrees well with the $0.7*m^{0.4}$ parameterization of the lower energy ISR results for masses up to about the $\phi$ mass.  Again, it should be noted that the ISR measurements only included masses below the $\Lambda$.  The strong trend of the $\langle p_\text{t} \rangle$ with particle mass could be an indication of the increasing importance of the mini-jet contribution to the $p_\text{t}$ spectra at RHIC energies as mentioned in \cite{Dumitru99}.

The transverse mass spectra from the $p+p$ collisions are seen to obey an approximate $m_\text{t}$--scaling in the low $p_\text{t}$ region but the greater range of the TOF spectra show the scaling breaks down above $\sim$1 GeV in $m_\text{t}$ for mesons but seems to hold slightly further out in $p_\text{t}$ for baryons.  The Au+Au data show a different behavior from the $p+p$ in that the radial flow, which is believed to be present in the Au+Au system, depletes the low-$p_\text{t}$ region producing a flatter, hotter spectrum.  The low-$p_\text{t}$ region then cannot be used for the scaling normalization, in effect breaking the approximate scaling observed in $p+p$.

\section{Acknowledgements}
We wish to thank the RHIC Operations Group and the RHIC Computing Facility
at Brookhaven National Laboratory, and the National Energy Research
Scientific Computing Center at Lawrence Berkeley National Laboratory
for their support. This work was supported by the Division of Nuclear
Physics and the Division of High Energy Physics of the Office of Science of
the U.S. Department of Energy, the United States National Science Foundation,
the Bundesministerium fuer Bildung und Forschung of Germany,
the Institut National de la Physique Nucleaire et de la Physique
des Particules of France, the United Kingdom Engineering and Physical
Sciences Research Council, Fundacao de Amparo a Pesquisa do Estado de Sao
Paulo, Brazil, the Russian Ministry of Science and Technology, the
Ministry of Education of China, the National Natural Science Foundation
of China, and the Swiss National Science Foundation.  This work has been
partially supported by NSF grant PHY-03-11859.


\begin{thebibliography}{99}
\bibitem{Rafelski82} J. Rafelski and B. Mueller, Phys. Rev. Lett. \textbf{48} (1982) 1066.
\bibitem{Bocquet96} G. Bocquet et al., Phys. Lett. \textbf{B366} (1996) 441.
\bibitem{BSCref1} B. Alper et al. Nucl. Phys. \textbf{B100} (1975) 237.
\bibitem{ISRparam} M. Bourquin and J. M. Gaillard, Nucl. Phys. \textbf{B114} (1976) 334.
\bibitem{Dumitru99} A. Dumitru et al.,  Phys. Lett. \textbf{B446} (1999) 326.
\bibitem{Gatoff92} G. Gatoff and C. Y. Wong,  Phys. Rev. \textbf{D46} (1992) 997.
\bibitem{McLerran94} L. McLerran and R. Venugopalan,  Phys. Rev. \textbf{D49} (1994) 2233
\bibitem{McLerran01} L. McLerran and J. Schaffner-Bielich,  Phys. Lett. \textbf{B514} (2001) 29
\bibitem{Bielich02} J. Schaffner-Bielich et al., Nucl. Phys. \textbf{A705} (2002) 494.
\bibitem{BielichXXX} J. Schaffner-Bielich et al., arXiv nucl-th/0202054
\bibitem{TOF} W. J. Llope et al., arXiv nucl-ex/0308022
\bibitem{NIM} Nucl. Instrum. Meth. A 499 (2003) 624
\bibitem{GEANT} http://wwwasd.web.cern.ch/wwwasd/geant/
\bibitem{JohnAndMark} J. Adams and M. Heinz, arXiv nucl-ex/0403020
\bibitem{STAR03} J. Adams et al., arXiv nucl-ex/0310004
\end{thebibliography}
\end{document}